# Predicting Physical and Physical-Chemical Properties of Molecular-Based Materials Using Computational Neural Networks

ANDREI A. GAKH, BOBBY G. SUMPTER and DONALD W. NOID

Chemical and Analytical Sciences Division, Oak Ridge National Laboratory,
Oak Ridge TN 37831-6197 USA

**GRAFICAL ABSTRACT**:

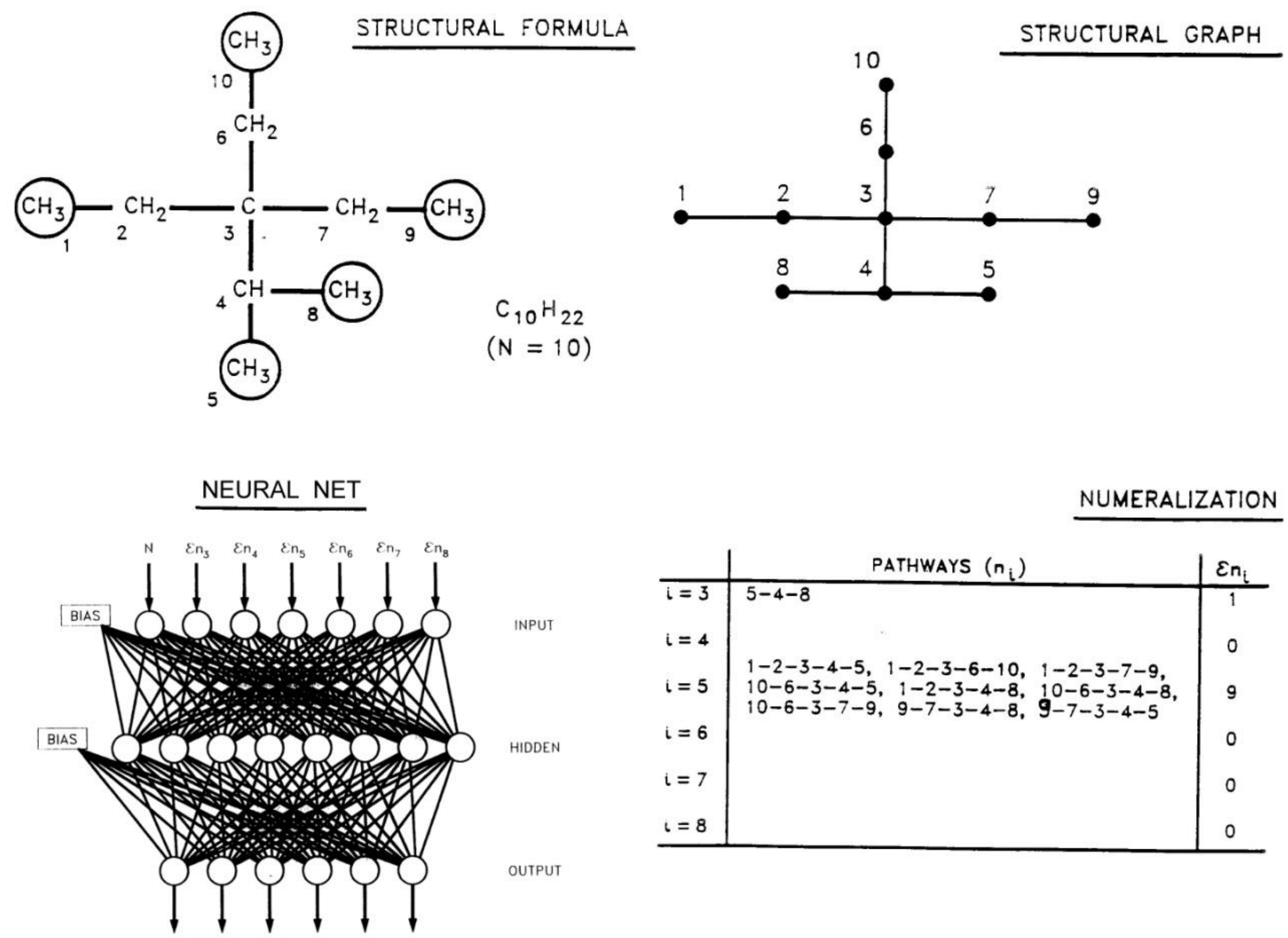

**ABSTRACT**: A computational scheme, which utilizes neural networks, was developed to predict properties of molecular-based materials from chemical structures. The method uses a set of simple algorithms to encode the structure and composition of organic molecules directly into numerical vectors, which is used as input for neural networks. Backpropagation type neural networks are then used to correlate these numeric inputs with a set of desired properties. Calculated results for a series of hydrocarbons, hydrofluorocarbons, and crown ethers demonstrate average accuracies of 0.2-8.1% with maximum deviations of 16-20% for a broad range of thermodynamic, physical, and physical-chemical characteristics (heat capacity, enthalpy, heat of evaporation, boiling point, density, refractive index, stability constants, etc.). In addition, a number of physical and mechanical properties were estimated for polymeric materials and compared with regression analysis. Based on the neural network capabilities of formulating accurate quantitative structure property relationships, a technique called computational synthesis is suggested for performing materials design.

## 1. INTRODUCTION

A common goal in materials science is the determination of relationships between the structure (microscopic, mesoscopic, and macroscopic) of a material and its properties (chemical, physical, biological) [1]. Such Quantitative Structure/Property Relationships (QSPRs) are crucial for engineering materials that provide a predetermined set of properties [2]. Although a materials design problem may often involve simply choosing the best existing material for a given task, the range of applications of some of the more common materials have been nearly stretched to their limits. Those suitable for advanced applications can have considerably more complex chemical structures or can be exotic blends. Furthermore, in order for a given material to be used in technology, it must first satisfy a number of performance criteria, and these criteria are becoming more and more stringent each year.

A purely empirical materials design approach (based on previous performance data, modern materials or processing techniques are formulated, produced, and tested) often requires a substantial amount of R&D. Obviously the ability to predict materials properties prior to their synthesis and processing would be of tremendous value in optimizing their end products [3]. Unfortunately, such capabilities are difficult to achieve, due to the large number of possible structures for a given composition (isomers), in addition to different processing techniques and complex aging processes. Furthermore, extracting the fundamental features needed to make structure-property correlations from existing data is not a simple task. Variability in the methods/procedures used to measure properties and conditions, such as the age of the material, leads to data that are sparse and noisy (so-called "real-world" data). Modern data analysis and inference techniques promise to greatly facilitate the development of models capable of predicting the desired performance specifications [4-11]. However, most methods capable of inferring underlying models with any functional dependence are asymptotically optimal (valid for high-quality and large data sets) [9]. These methods are ill-posed for analyzing the "real-world" data. Thus it is prudent to understand which methods are most appropriate for the problem at hand.

The investigation and development of structure-property relationships based on existing "real world" data requires methods that make no assumptions about the model, and that are capable of dealing with multivariate and complex functions with possibilities of complicated forms of inter-variable correlation. Computational neural networks are one of several methods that are capable of meeting these requirements [11].

The use of neural networks of QSPR is somewhat different from other computational methods and has both advantages and disadvantages. The primary advantage is that the self-adjusting (backpropagation) algorithm does not require theoretical formulas or postulated models. The disadvantage is the necessity of presenting the structure and composition of a material into a special numeric form acceptable for neural network computing. In addition, a relatively large set of the structurally similar compounds with experimentally determined properties is needed for training [11].

In the present paper we discuss results of QSPR analysis of molecular-based materials obtained using backpropagation type neural networks. The essence of the computational method lies in the application of simplified procedures for generating numeric input based primarily on the structure and composition of organic molecules [12-15].

## 2. METHODS

### *Numeric Representation of Molecular Structure and Composition*

The numerical representation of structure and composition of organic molecules in a form suitable for a neural network remains a challenging problem [16]. A general approach calls for an algorithm applicable to the wide variety of molecules including organometallic compounds, complexes, polymers, and biomolecules. The complexity of the task implies that only sophisticated universal algorithm(s) can be used. However, much more simple methods can be successfully applied for a specific class of organic compounds. The best choice of a coding method will depend on similarity basis used for definition of a class.

In simple cases, such as in the case of non-cyclic saturated hydrocarbons, the class is defined as a set of compounds consisting of only $sp^3$ carbons and hydrogen connecting to each other by single bonds. It can be easily elucidated that these compounds should have a general formula $C_nH_{2n+2}$ and differ from each other only by the topology of carbon network and number of carbon atoms. Applications of topology methods for representation of structural features into numerical input for a neural network are straightforward and obvious.

A standard approach in this case is computation of structural graph invariants, i.e., topological indices [17]. In our work we utilized a simple "local sum" algorithm which was developed on the basis of Wiener-type topological indices [18]. The algorithm is based on the identification of all possible shortest bond distances (pathways) between terminal carbon atoms (in $CH_3$ groups) [12]. The next step is calculation of the numbers of the pathways with the same length "local sum". (As the length of the pathway we used a number of the carbon-carbon bonds between the atoms). For simplicity, we considered all the pathways with the same length as equal, and the pathways with the length of more than 8 bonds were omitted. With these simplifications the structural data part of inputs is reduced to only 5 digit numbers. Although these simplifications have led to the loss of the reverse uniqueness (we have obtained the same sets of "local sum" numbers for some hydrocarbons with the different structures when n > 10), these examples are rare. A sample of an alkane structural graph representation, "local sum" numbers calculations and final inputs generation is shown in Figure 1 [12]. In addition to structure-related data, inputs also contain the number of carbon atoms as a separate entity to represent the composition of a molecule.

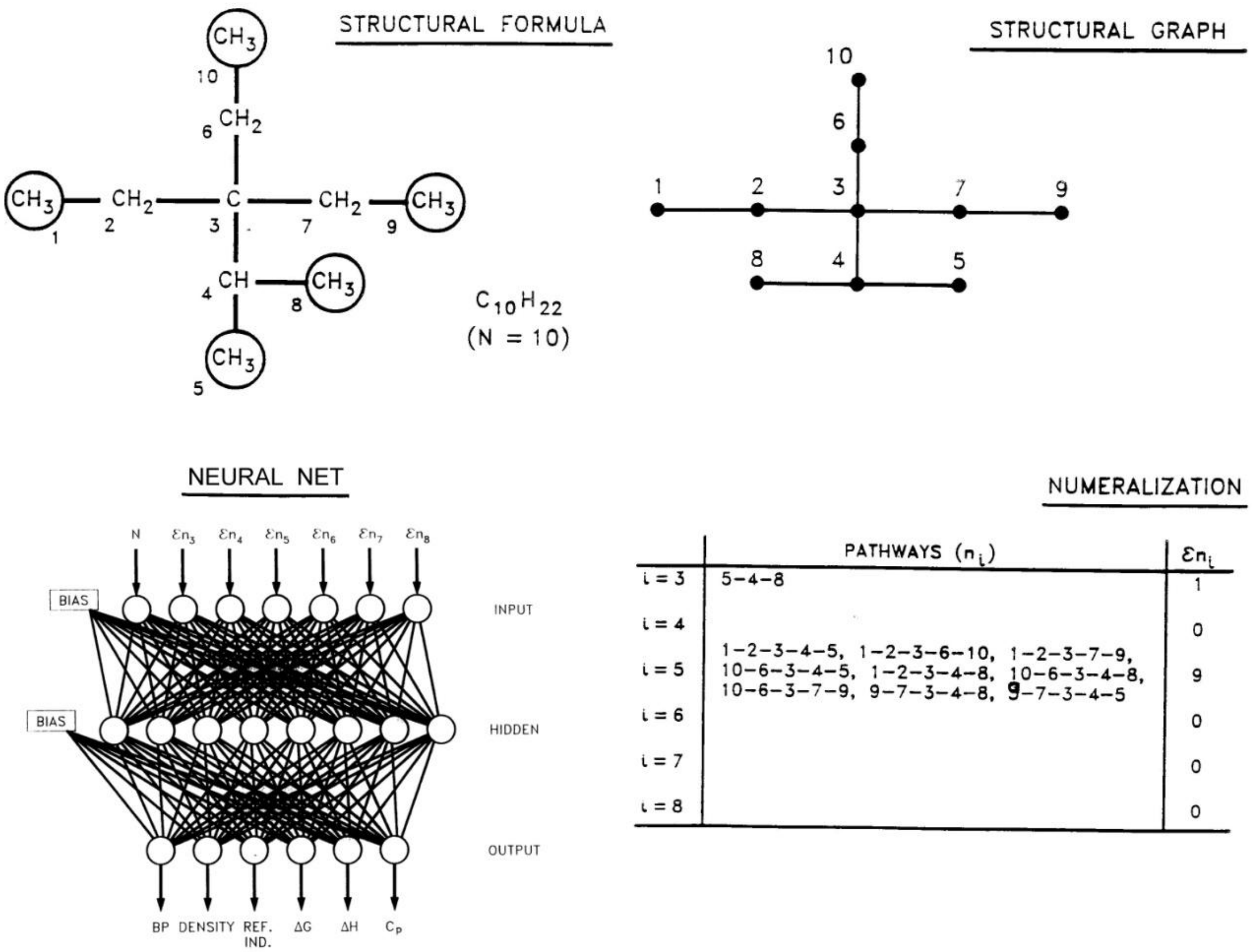

FIGURE 1 Sample generation of the input file, structural codes, and neural network configuration used for prediction of properties of hydrocarbons.

Aliphatic non-cyclic hydrocarbons is one of the simplest types of organic molecules, since they contain only two structural elements-hydrogen and $sp^3$ carbon. Addition of one more element, fluorine, complicates the molecular structure and composition coding. In addition to carbon skeleton isomers, hydrofluorocarbons (HFCs) also have fluorine position isomers. In this case we have to separate the hydrofluorocarbon class into subsets of hydrofluoromethanes, -ethanes, -propanes, etc. Since each of the subsets has a uniform carbon skeleton, they are ideal objects for another simple coding technique, which generates numeric input vectors directly from structural formulas (see Figure 2) [13]. This coding method is also capable of some generalization, thus allowing us to utilize one neural network for three subsets of compounds - hydrofluoroethanes, -propanes and -butanes. To make all numeric inputs compatible (8 digits), we added 00 or 00 00 in hydrofluoropropanes and -ethanes numeric vectors. These double-zero values indicated that the corresponding structural elements were missing. Since our scheme cannot distinguish properties of diastereomers, their physical data were averaged.

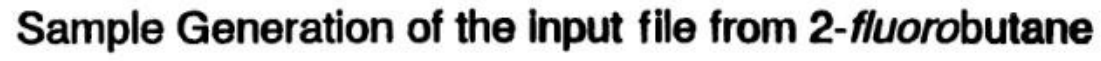

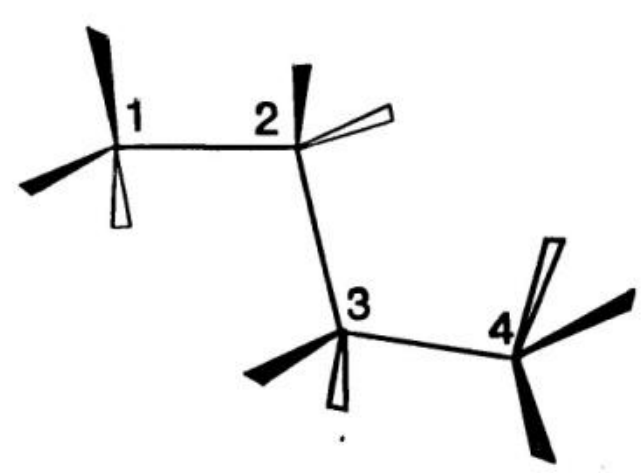

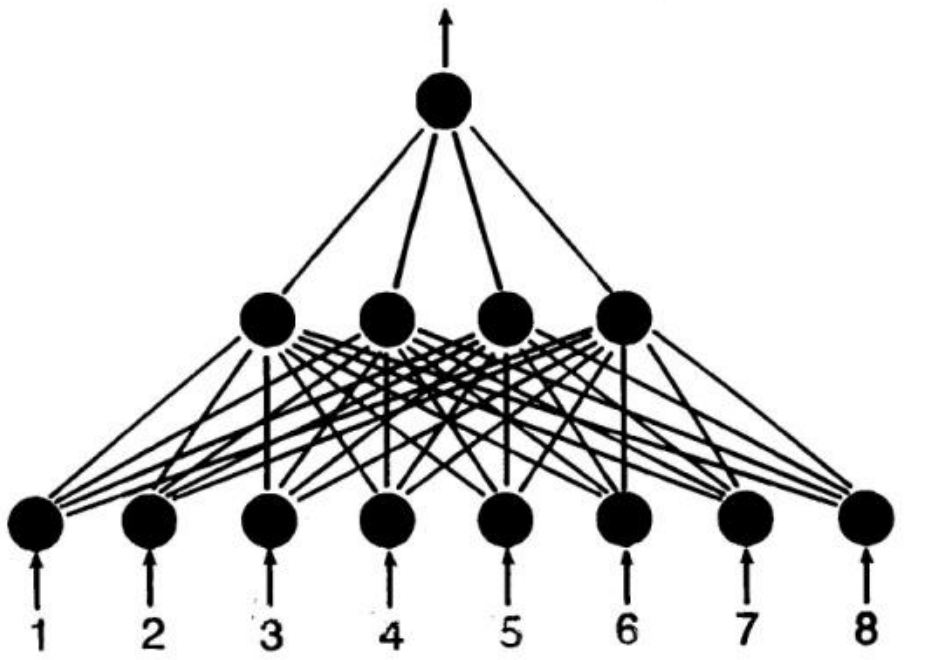

FIGURE 2 Sample generation of the input file and neural network configuration used for prediction of properties of HFCs. Bias connections in the neural network schematic were omitted for clarity.

Even more complex organic molecules can still be easily coded for neural network inputs, providing they have similar (or repeated) structural elements. In this case, the numeric vectors can be derived directly from the systematic names of the organic compounds [14]. A classical example of such a class of organic compounds is crown ethers.

Simple crown ethers are cyclic oligomers containing donor atoms (usually oxygen or nitrogen) in repeating structural units. For our initial study, we chose monocyclic crown ethers consisting only of aliphatic 1,2-dioxyethylene and aromatic 1,2-dioxyphenylene fragments. Since the exact positions of these units are usually omitted in the chemical names of these compounds, our computations include only ring size, overall number of oxygen atoms, and number of aromatic rings. Thus, for example, the input vector of benzo-18-crown-6 (B18C6) is 1 18 6, where the first digit represents the number of benzene rings (benzo in the name), the second digit represents ring size (-18- in the name), and the final digit represents the number of oxygen atoms in the ring. Accordingly, for dibenzo-21-crown-7 ($B_2$18C7) the numeric input vector will be 2 21 7. An example of a sample numeric input vector preparation is presented in Figure 3 [14]. A more general treatment of the method for numerically representing chemical structure based on their systematic names is given in the appendix.

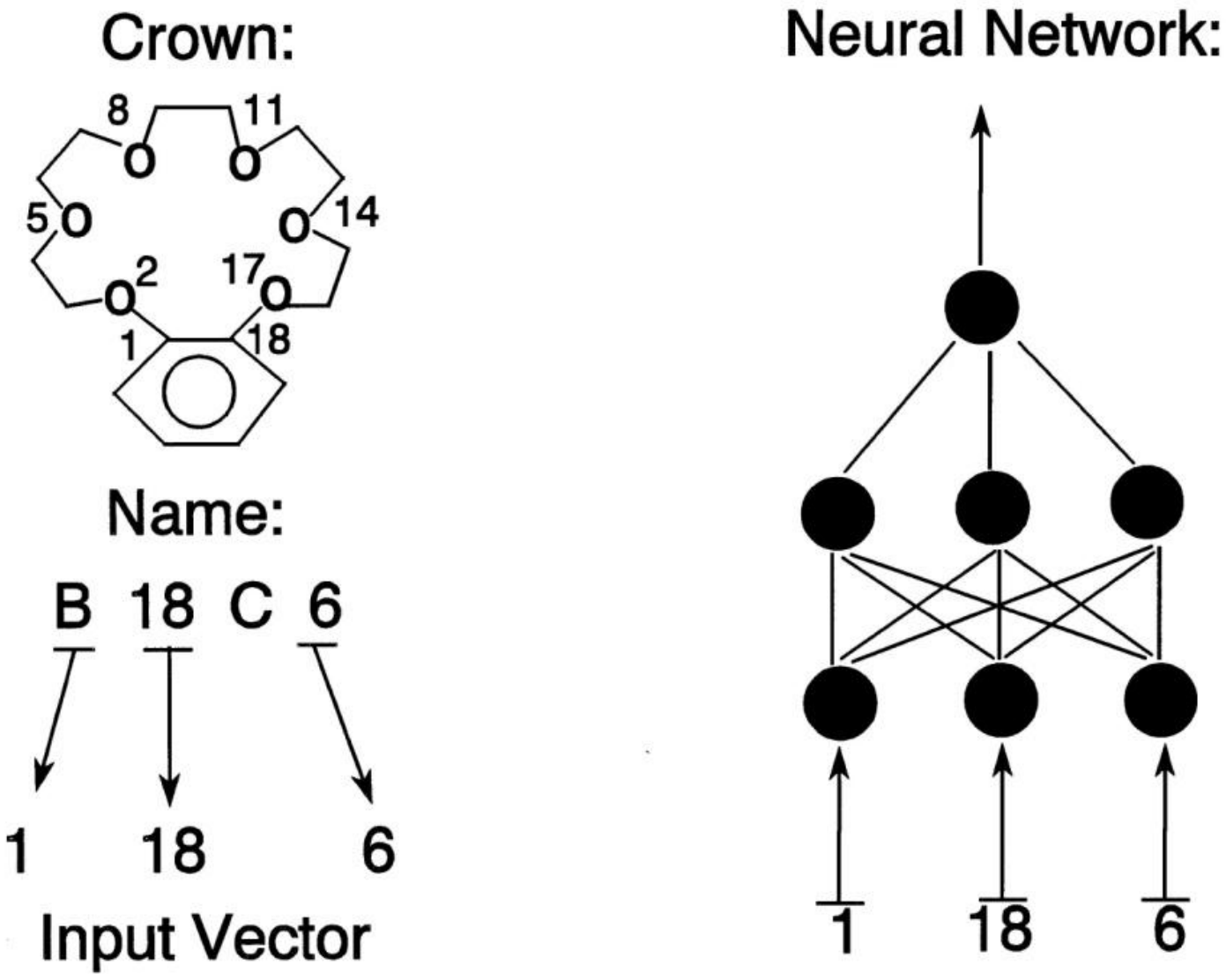

FIGURE 3 Sample generation of the input file and neural network configuration used for predicting the stability of crown ethers complexes. Bias connections in the neural network schematic were omitted for clarity.

***Backpropagation Type Computational Neural Networks***

Computational neural networks (CNNs) are model-free estimators that have exceptional ability for performing multidimensional, nonlinear vector mappings [19-22]. In general, CNNs have several essential constructs that define their operation: nodes (simple processing units), transfer functions (generally nonlinear and bounded functions), connection weights, and a learning algorithm. One possible arrangement of the nodes is an architecture that describes a multilayer feedforward network: a set of nodes placed into two or more layers. There is an input layer and an output layer, each consisting of at least one node. The nodes in the input layer do not perform any actual processing, but serve only to distribute the input to the next layer. There are usually one or more hidden layers (layers of nodes between the input and output). The term *feedforward* means that the inputs to the nodes in each layer comes exclusively from the outputs of nodes in the previous layer, and the outputs from these nodes pass to nodes in the following layer. Each node in the network has a number of weighted connections to other individual nodes and an input signal is propagated through the system until it emerges as a network output. An optimization procedure that adjusts the weights connecting the nodes in order to minimize the difference between the output and the target (the desired result) is called the learning algorithm. Backpropagation was the first practical method for training multilayer feedforward networks and is still the most popular learning algorithm.

The backpropagation algorithm is an example of supervised learning (training with a teacher, that is, with known answers for representative examples). This algorithm adjusts the weights based on a gradient descent minimization of an error function (usually the sum of squared errors). The goal is to "teach" the network to associate specific output to each of several inputs. Having learned the fundamental relationship(s) between inputs and outputs, the neural network should then be able to produce reasonable output for unknown input (called generalization). In practice, this type of iterative approach can often take a relatively large number of epochs (cycles through the whole data set) before a reasonable error is reached. Fortunately there is a number of methods that can help alleviate this pathology, although each modification often comes with a number of new problems. For example, we have had very good success with using the Levenberg-Marquardt compromise to Newton's method for optimization problems [23, 24]. This method basically interpolates between gradient descent and Gauss-Newton methods, depending on the distance away from the minimum (gradient descent is used for minimizing the error for positions that are far away from a minimum and Gauss-Newton is used for those close to a minimum). This approach is quite practical (for < 100 variables) since it is well known that second-order or quasi-second-order methods converge much faster than gradient descent close to a minimum (the error surface is approximately quadratic), but are slower when the error surface is not parabolic (far away from the minimum). However, if the number of weights for the optimal neural network architecture is greater than 100, the Levenberg-Maquardt method becomes very inefficient. In this case we have found that the Polak-Ribiére conjugate gradients method with Powell restarts provides an excellent optimization method for neural networks [23]. Addition of a stochastic perturbation term to this method, much in the same spirit as Langevin dynamics, can also give improved performance [25].

One final note, rather using the standard backpropagation or one of its many variants, it is sometimes prudent to carry out all calculations in the complex domain [26]. This type of approach is usually referred to as *complex backpropagation* and can actually lead to some surprising advantages. Training speed and reliability usually increase dramatically, and generalization quality is almost always superior. The implementation is a minor modification of the equations used in backpropagation: there will be a real part (Re) and an imaginary part (Im) to the input, output, bias, transfer function, and connection weights. Complex domain neural networks are appropriate, primarily when the data intrinsically occur in pairs of numbers.

For many users, the applications of backpropagation neural networks are greatly facilitated by the availability of commercial software packages. Most of the computation in this paper can be performed on a PC, using one of the most popular packages developed by NeuralWare (*Neural Works Professional II PLUS*) [27]. This package includes a comprehensive manual and many tools, which allow real-time control of neural network performance.

Four major parameters play an important role for the success of neural networks computing using this package: transfer function, momentum, learning rate, and number of neurons in hidden layer. Although the role of these parameters in neural networks computing has already been well documented [11], we would like to emphasis some empirical observations concerning *Neural Works Professional II PLUS* software package.

In a large number of our studies the influence of the transfer function was found to be critical. Unfortunately, there are no clear rules on what function is the best for a specific task, but some empirical guidelines can be suggested. *Neural Works Professional II PLUS* currently has an option to use three popular transfer functions: logistic, hyperbolic tangent, and sinus. We have found that the right choice of transfer function is governed mostly by the character of input/output relationships in a data set. When such relationships are relatively smooth (e.g., evaluation of density of hydrofluorocarbons based on their composition and structure), the logistic function is a function of choice. In more complicated situation, such as in case of boiling points, the hyperbolic tangent is better. The optimal values of other two important parameters, learning rate and momentum, mostly depend on the choice of the transfer function. Usually the logistic function calls for higher learning rates and momentum than the hyperbolic tangent. We recommend starting training with lower parameters, and slowly increasing them to the point when the neural networks become unstable (strong fluctuations of error, bad distribution of connecting weights on weights histogram) [27]. The best performance is usually achieved near the critical point.

### *Data Preprocessing*

The performance of a CNN is strongly dependent upon the patterns used to train it (training set). The training set must provide a full and accurate representation of the problem domain if the network is to meet expectations. Several goals should be satisfied: (1) every class must be represented; (2) within each class, statistical variation must be adequately represented. If advanced knowledge on the behavior of the data (for example, if some patterns are easier to classify than others) is available, shaping the data by weighing patterns (overrepresentation) can often be advantageous. These guidelines provide a reasonable way to collect a training set provided that a sufficient amount of numerical data is available.

Once a suitable training set has been collected, preprocessing the input data often plays a very important role in the CNN's ability to carry out the desired task. Proper scaling or normalization of the data can help improve the performance. Scaling the data to values that fall in the range of the bounds of the transfer is essential when using output nodes with a bounded transfer function. In many commercial software products such as *NeuralWorks Professional II/PLUS* scaling is performed automatically.

Supervised neural networks require a training and a testing data set. Unbiased selection of these sets from the available data can be achieved by using the "bootstrap" resampling method [28]. This method, founded in statistics, is a powerful procedure for determining the best estimator for small data sets. "Bootstrap" resampling is basically random sampling with replacement: a data set of N examples is randomly sampled N times to create a new data set with N examples. The new data set will have the possibility of sample repetition and a test set can be generated by comparing the new data set with the original one and selecting those examples that are unique. Thus the "bootstrap" method can be used to produce a number of different training and test sets and the error estimate is taken as the average performance on the data set ensembles. In general it has been found that about 200 iterations of the bootstrap estimates are needed to obtain a good representation.

"Bootstrap" resampling is only one of several methods that provide desirable properties in error rate estimates. Another method that works very well for neural network training and testing is the "jackknife" method [28]. The general procedure is to take 1 (it also can be generalized to k samples) sample out of the available data as a test set and train on the remaining ones. This is repeated for all of the samples, producing an ensemble that is the same size as the original data set. This procedure can be used in an identical fashion as described for the "bootstrap", but requires fewer runs.

By using statistical resampling ("bootstrap" or "jackknife"), a CNN can be trained on an ensemble of different training sets and the performance (cross-validation) evaluated on the compliment testing sets to determine a measure of the true performance (the average over the ensemble). Some added benefits on error performance can also be obtained by averaging the network's error over an ensemble of initial connection weights (in general, one set of connection weights will lead to convergence to a different local minimum than another set). This approach, called the *ensemble average method* [29], not only gives a better estimate of the true error (unbiased) but also reduces the size of the neural network (number of hidden nodes) over that required for any single training/testing computation and tends to smooth out (by averaging in function space instead of parameter space) the effects of overfitting (variance reduction). Thus, an optimal CNN model for a given set of data can be obtained by performing ensemble averages over initial connection weights and training/test sets. This procedure is obviously more computationally demanding than one consisting of a single run, but adequately makes up for this pathology, since it provides an optimal and unbiased representation of the performance capabilities of a CNN. One final comment - the method of stopped training, the error measure is minimized with respect to performance on cross- validation test sets, is used to determine when a particular neural network architecture has reached its optimum performance (avoids overtraining).

## 3. RESULTS AND DISCUSSION

One of the most popular concepts in chemical, biological, environmental, and materials science is that there exist strong correlations between the properties of a molecule and its chemical structure. Structure activity and structure-property relationships have been pursued for many years under the titles, quantitative structure-activity relationships (QSAR), quantitative structure-property relationships (QSPR), and quantitative structure (you fill in the name) relationships (QSXR). Much of the research in this field has been motivated by the desire to design new and better compounds for specific applications, particularly in pharmaceutical and materials sciences. Recently, notable improvements have been found by using CNNs, which provide a flexible platform capable of determining the complex correlations that exist between structure and activity/property. Here we present some of the results obtained by this method [12-15].

**Hydrocarbons [12]**

Backpropagation-type neural networks described above have been applied to learn and predict the structure - property relationships for the set of data for 134 hydrocarbons, $C_6$-$C_{10}$ [30]. The network architecture which worked best for us (see Figure 1) consisted of two layers (one hidden and one output layer, not counting the input layer) with a topology of 7 input nodes, 8 hidden nodes, and 6 output nodes. Bias nodes and short-cut connections were also included, thus giving a total of 166 connection weights. Training of the neural network made use of momentum (0.9) and an adaptive learning rate, based on the gradient. The final architecture was determined from a number of experiments, which were aimed at finding a relatively small network capable of predicting the training data consisting of 109 examples and of making generalizations to a test set consisting of 25 examples. To insure that the network did not overfit the training data (memorize it), the jackknife method, combined with cross-validation, was employed. Absence of systematic errors and a relatively low level of both average deviation (1.3-2.7%) and maximum deviation (14.6%) were unmistakable signs that the chosen neural network can correlate structural and stoichiometric parameters with the physical properties for a representative set of hydrocarbons.

The trained neural network was then used as computational tool for prediction of the properties of a set of 25 hydrocarbons with diverse structures and molecular composition. The results of the predictions are presented in Table I. As can be seen, the neural network predicts the desired properties with an average error of less than 2% and maximum deviation of less than 12%. Average deviations of calculated data from the experimental results are presented as a histogram in Figure 4. The comparison with other known computational methods, based on the structural graph representation, indicates that our method is among the most accurate ones [31].

The analysis of the data obtained shows that very different physical properties (e.g., boiling point and density) of the compounds could be predicted with low average errors (0.2-1.2%). This implies that our computational model correctly represents structure-property relationships in these classes of compounds, and that our numerical inputs correspond to structural features of the hydrocarbons responsible for their physical characteristics. It is also worth noting that the average errors in both the training and prediction sets are close, which suggests that, after training, the neural network represents a correct "model" of the structure-property relationships.

Analysis of the results of the neural network computations revealed that the best predictions were achieved for refractive index and density of the hydrocarbons. These parameters are not as sensitive to the specific, unique interactions in the molecules. Our computational scheme averages the structural features in the path numbers and allows effective approximation of such parameters. The same arguments can be applied to heat capacities, which were predicted with an accuracy as high as 99.1%. The predictions of boiling points were less accurate (98.8%) and of thermodynamic functions such as Gibbs energy and enthalpy of formation, somewhat worse (98.6-98.5%). The main reason for this is the known strong dependence of these properties on noncovalent interactions. These noncovalent interactions were not elucidated in a full range by the chosen set of descriptors. As a result, thermodynamic properties of sterically hindered and strained molecules were predicted with less accuracy than non-hindered ones (2-3% on average, compared with 1-2%). Addition of a parameter which could reflect such noncovalent features of molecules might be of help to improve the accuracy in these cases.

TABLE I Average errors of property predictions using computer neural networks *

| Class of Compounds | Property | | | | | | |
|---|---|---|---|---|---|---|---|
| | Boiling Point | Density, at 25°C | Heat Capacity | Heat of Vaporization | Refractive Index | Enthalpy at 300°K | Stability Constant |
| Hydrocarbons | 1.8°C<br>1.2% | 4.4 kg·m$^{-3}$<br>0.6% | 1.8 J/M·deg<br>0.9% | —<br>— | 0.003<br>0.2% | 0.5 kJ·M$^{-1}$<br>1.4% | —<br>— |
| Fluorohydro-carbons | 10.8°C<br>— | 30 kg·m$^{-3}$<br>2,7% | —<br>— | 1.1 kJ·M$^{-1}$<br>4.5% | —<br>— | —<br>— | —<br>— |
| Crown Ethers | —<br>— | —<br>— | —<br>— | —<br>— | —<br>— | —<br>— | 0.34 LogK<br>8.1% |

* The error measures given in Table I are defined as follows: (error) = Σ[abs ($T=P$)]/$n$, where $T$ is the known answer, $P$ is the prediction and $n$ is number of examples: (%error) = Σ[abs ($T-P$)/$T$ * 100]/$n$.

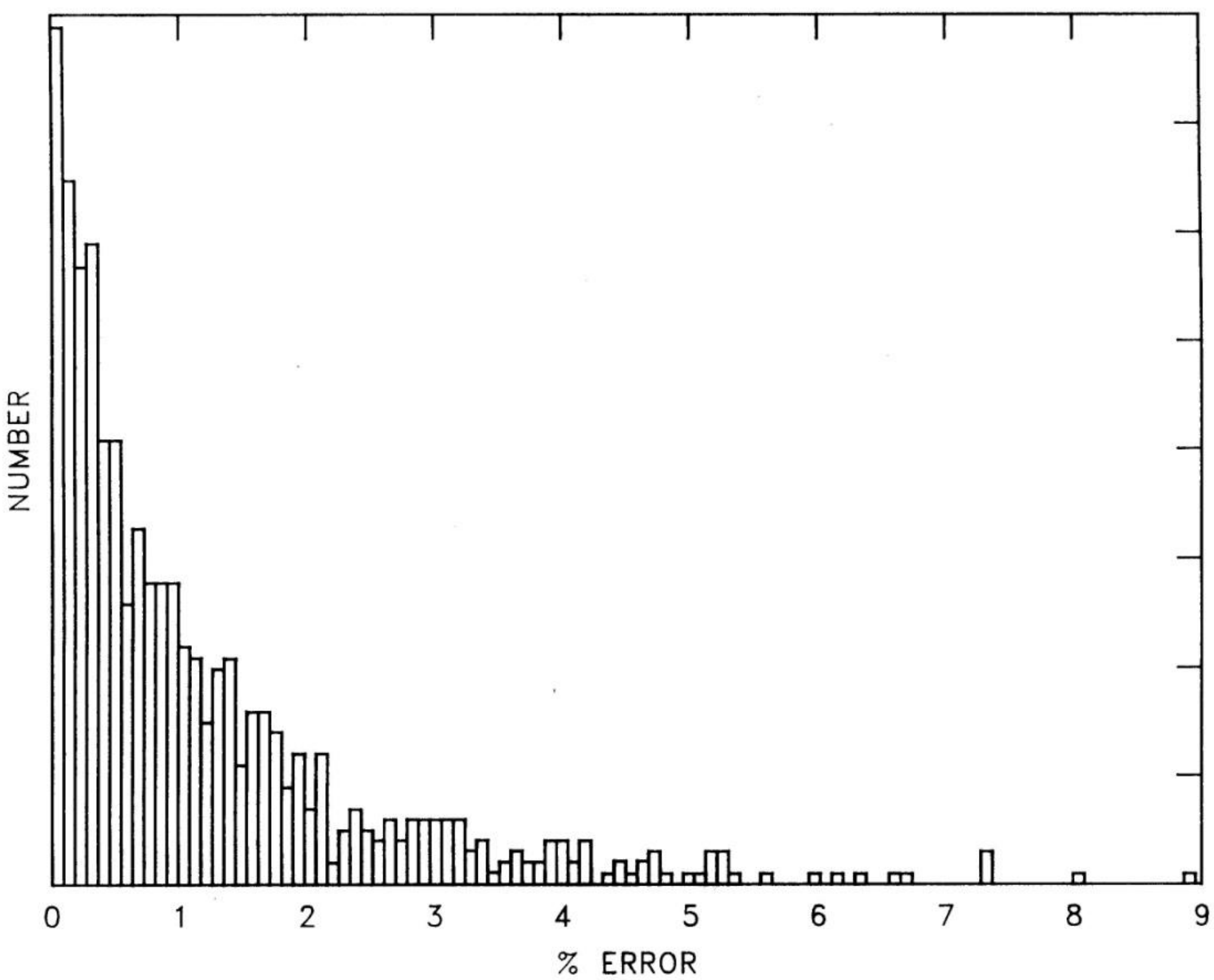

FIGURE 4 Histogram of the absolute % error from the neural network calculations of hydrocarbon properties, abs[(Experiment - network)/Experiment]∗100.

There is also a noticeable difference between the accuracies of the predictions of the properties of low ($C_6$) and high ($C_{10}$) hydrocarbons: the average error being significantly higher for hexanes (2.5%) than for decanes (0.7%). The same trend was also observed for the training set. We suggest that this is mainly an effect of an inadequate training set: there are only 5 isomers of hexane, compared with more than 70 structural isomers of decane.

Other important information was obtained from a sensitivity analysis of the neural network. Data demonstrated that the importance of the input values quickly diminishes with the increasing length of the pathways. This indicates that the structurally important information is concentrated is short-range (2-4 bonds) pathways. We assume that for more complex molecules containing heteroatoms, the length of the "important" pathways may be limited to 5 bonds, thus reducing the structure-related input to only 4 digits.

**Hydrofluorocarbons [13]**

Among the most essential properties of industrially important hydrofluorocarbons are boiling point, heat of evaporation, density, critical temperature, atmospheric lifetime, and ozone-depletion potential. Using our technique, we have tried to predict some of the above properties. To ensure better flexibility, each of the properties was calculated independently using a separate neural network with individual sets of parameters such as number of neurons in hidden layer, learning rate, transfer function, and momentum. Different sources of physical data, including reference materials, primary publications as well as computer databases, were used for training in this study [32-36]. In the cases of ambiguous data, the preference was made to the most recent publications.

There are several reports on predicting boiling points of CFCs and HFCs. The most recent one is based on data for 256 CFCs, mostly propanes, ethanes, and methanes, and includes equations found by regression analysis which correlates boiling point with a set of topological indices and stoichiometric data [37]. However, the best accuracy was achieved by a modified "boiling point numbers" scheme [38], which was initially developed to establish relationships between stoichiometry and boiling point. The modification included some additional rules, which were established in the series of halogenated ethanes on the basis of structural fragments analysis [38].

Our neural network performance data are presented in Table I and Figure 5. These results were achieved by a neural network having 6 neurons in hidden layer (hyperbolic tangent) after 128,032 epochs [27]. The average error of the test set was 10.8°C, which is noticeably worse than what has been reached by regression analysis (5-6°C) [37, 38]. The main reason for this is that utilization of the generalized input file (ethane, propane, butane derivatives all together), and the well known dependence of the boiling points of HFCS on dipolar interactions [38]. The last parameter is the most important one since we have already shown that a similar approach for predicting boiling points of hydrocarbons gives an average error as low as 2-3°C [12]. It appears that the neural network experienced difficulties in recognition of these dipole-dipole interactions (not counted directly in our scheme). The overall performance was sufficiently better (average error 4.7°C) when we excluded the extremes and used a limited temperature range of -40 to +40°C (Figure 5).

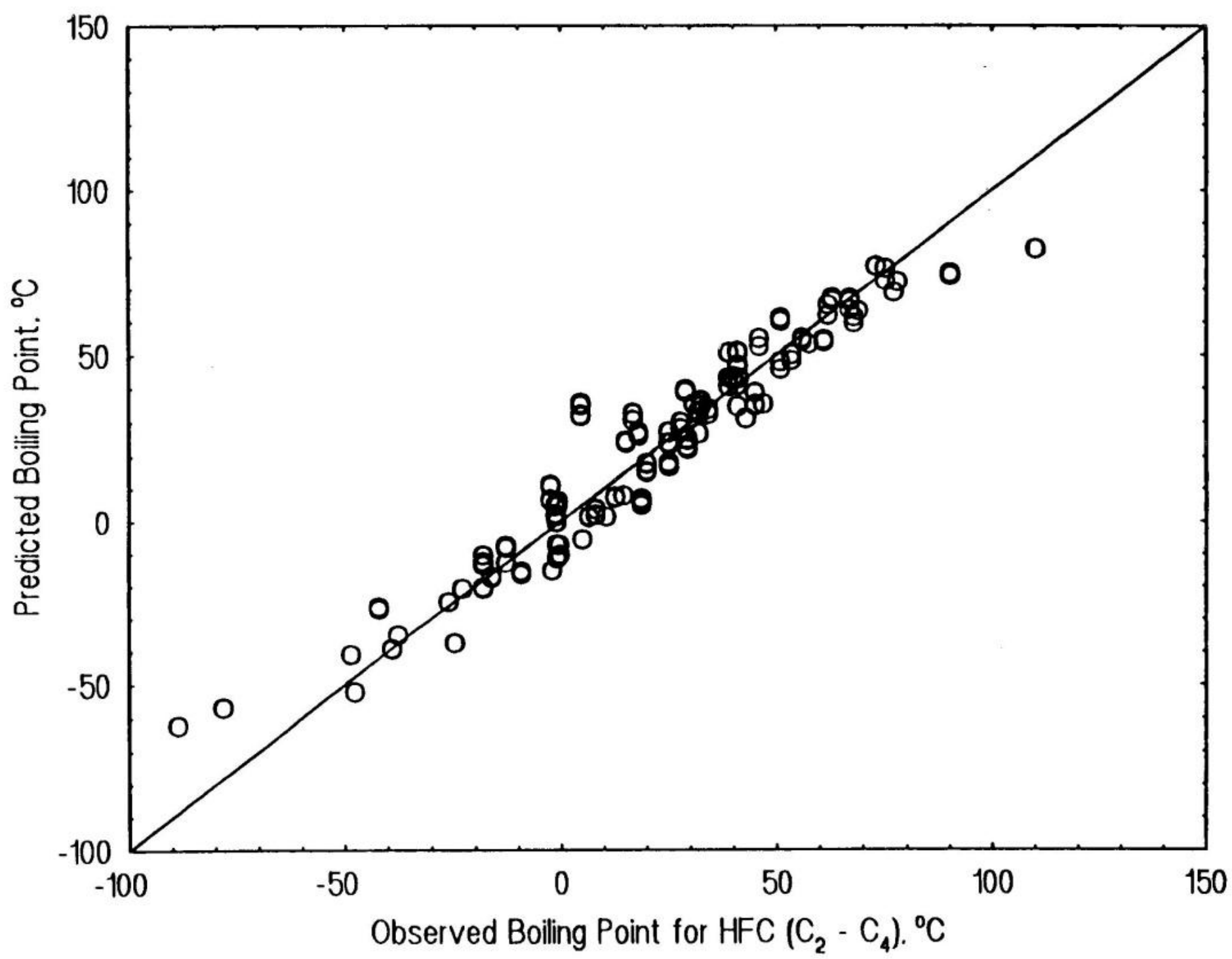

FIGURE 5 Experimental versus neural network predicted values of the boiling point of HFCs.

The heat of vaporization represents another important parameter for HFCs application, especially in refrigeration and air conditioning systems. Since this parameter is much less sensitive to the specific features of the molecule, the results are usually good, even in the case of HFCs (see discussion above, Hydrocarbons section). The optimized architecture contained one hidden layer (4 neurons, sigmoid function), and the training cycle was completed after 44,467 epochs [27]. The results are presented in Figure 6 and Table I.

Density is also important property of HFCs, and it appears to be the best parameter for the predictions based on neural network computations. As we have already shown above, density of alkanes could be predicted from their structures with an average accuracy better than 99%. Density of HFCs can be predicted with lower, but still very good accuracy (97.3% or ± 0.03 g/cm$^3$), and is satisfactory for most applications. These results were achieved with one hidden layer containing 2 neurons (sigmoid function) after 30,021 epochs [27]. The results are presented in Figure 7 and Table I.

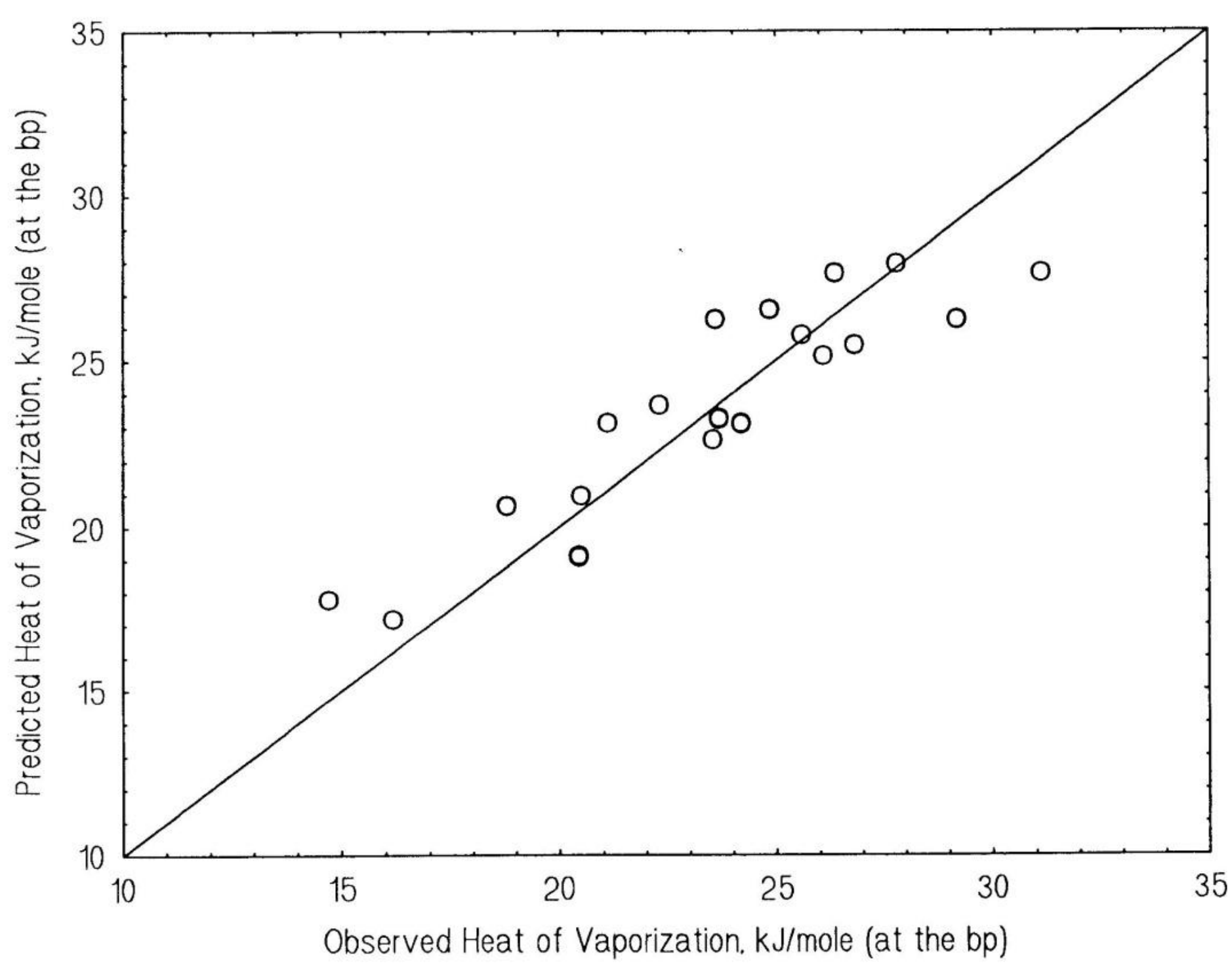

FIGURE 6 Experimental versus neural network predicted values of the heat of vaporization of HFCs.

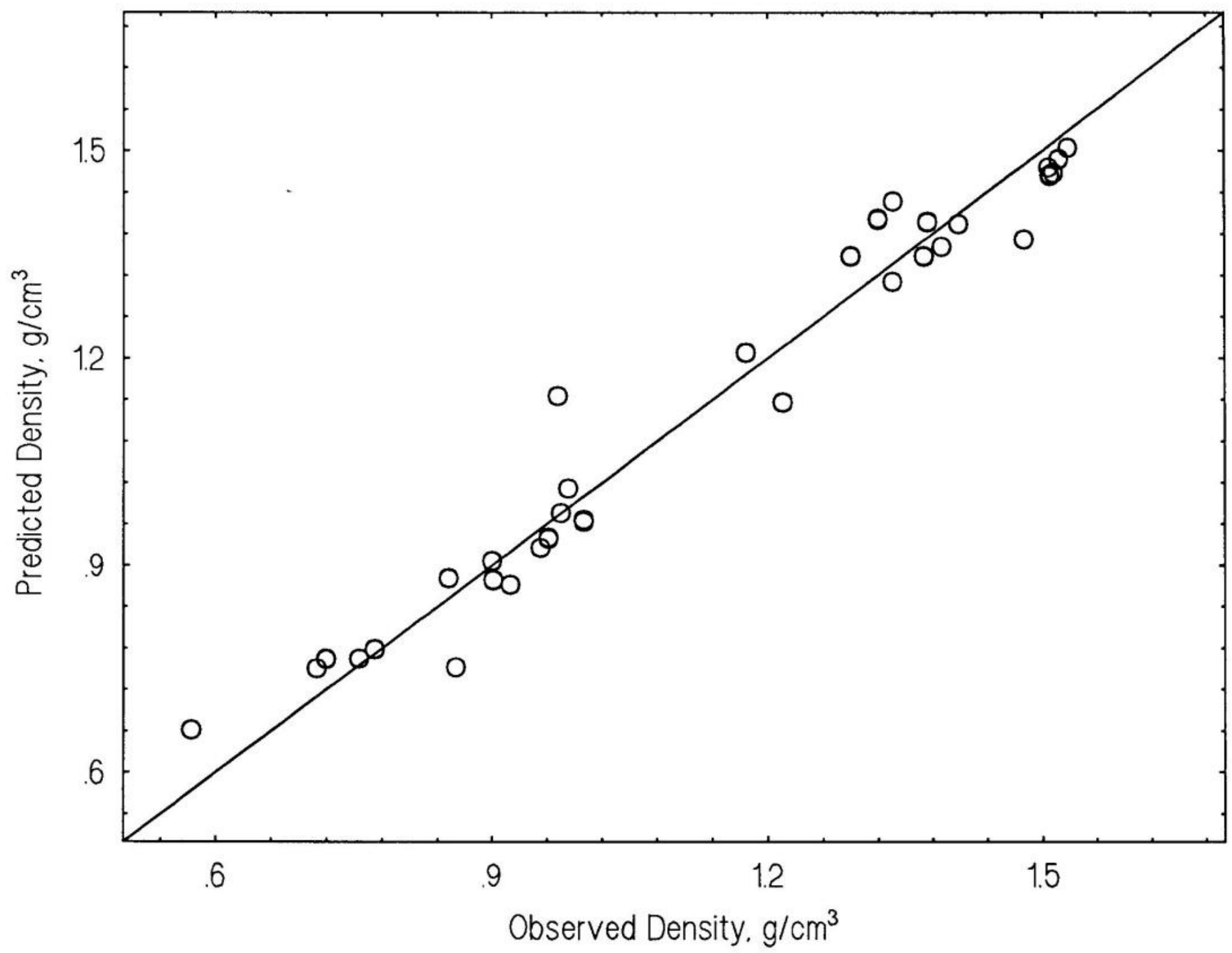

FIGURE 7 Experimental versus neural network predicted values of the liquid density of HFCs.

**Crown Ethers [14]**

A computational scheme discussed above was used for the prediction of complexation constants of simple crown ethers with alkali metal cations in a single solvent (methanol) at 25°C. Only the chemical information contained in the name of the crown ether was employed.

Stability constants of various crown ether complexes with alkali metal cations (output data for training and testing) were used as reported previously without any preselection [39, 40]. In the case of conflicting data, all sets of available constants were taken as input. Since the temperature and solvent also affect the stability constants, we used only data obtained at standard temperature (25°C) and in one solvent (methanol).

The optimized neural network (optimization procedure included leave-k-out and other previously mentioned methods) consists of 3 input neurons, one hidden layer (3 neurons), and one output neuron (see Figure 3). For simplicity, bias was omitted on the picture.

Our neural network performance data are presented in Figure 8. These results were achieved by the above neural network having 7 neurons in all layers (linear function for input layer of 3 neurons, hyperbolic tangent for the subsequent layers) after 86,000-123,000 epochs. The average error of the training sets was (± 0.27 Log K units or 6.4%. The average error of the testing sets was slightly higher (± 0.34 Log K units, or 8.1%), an indication that the neural networks are successfully recognizing structure property relationships, not simply memorizing the training set.

Successful construction of neural networks that represent computer models of complexation phenomena allowed us to investigate some tendencies in the stability constants of alkali metal-crown ether complexes. For that purpose we prepared artificial input files for crown ethers with the general formula $(\text{-CH}_2\text{CH}_2\text{O-})_n$, and calculated Log K using previously trained neural networks. The results were plotted for the three alkali metal cations, Na, K, and Cs, versus ***n***, in Figure 9.

The depicted results are in general agreement with the well-known correlations between the ring size and the ability of crown ethers to complex alkali metal cations [39-43]. The main feature of these correlations is the existence of maxima on the ring size-stability constant curve, which represent the structure of the crown ether best suited for the complexation of the specific metal cation. For sodium and potassium, the optimum binding is obtained with 18C6; smaller and larger crown ethers bind more weakly. For Cs, a maximum is reached with 24C8; however, the decrease in binding strength with increasing ring size is small.

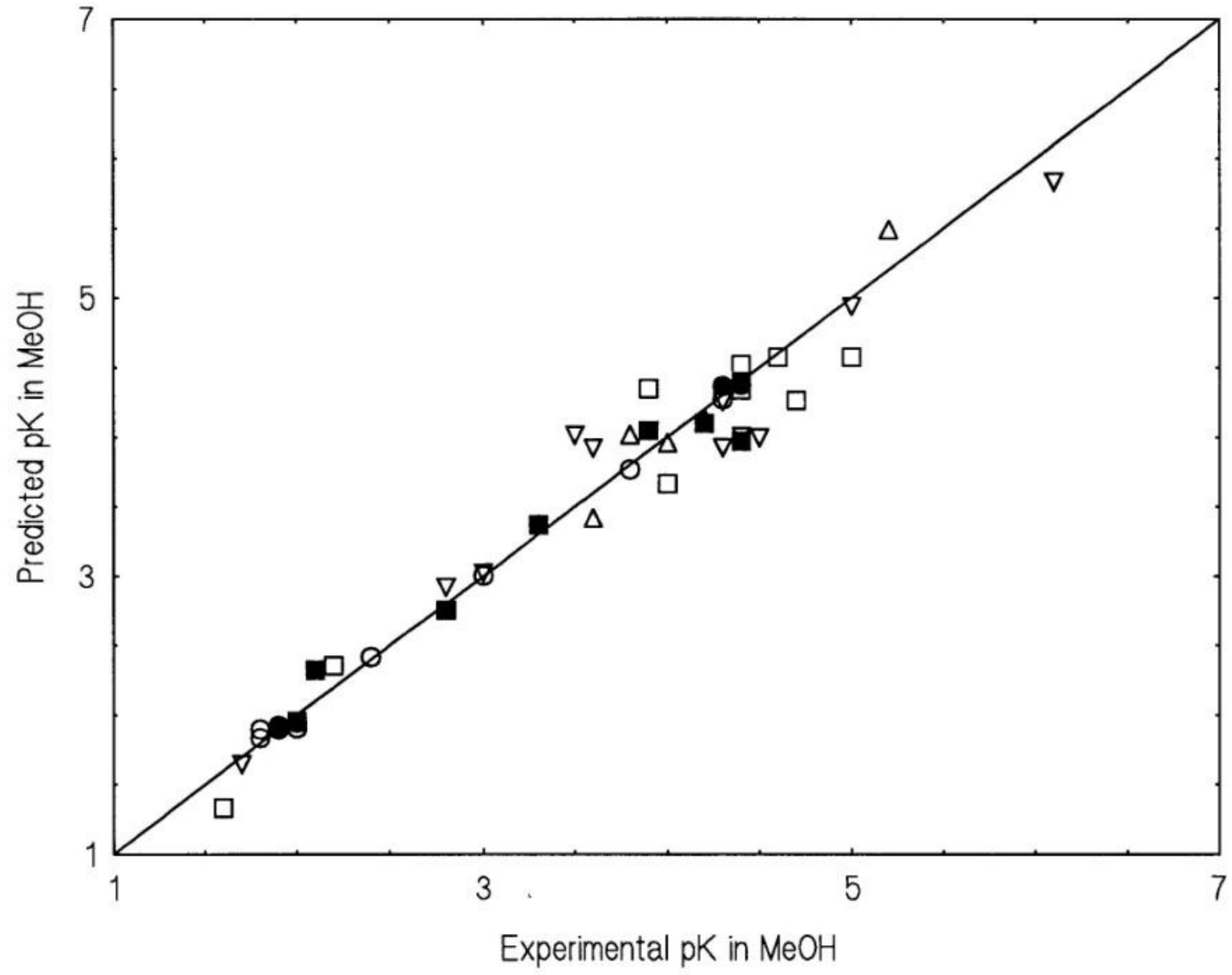

FIGURE 8 Experimental versus neural network predicted values of the stability of crown ethers complexes: a) $Na^+$: learning (training) set - open circles; testing (validation) set - filled circles, b) $K^+$: learning (training) set - open triangles with down vertex; testing (validation) set - open triangles with up vertex, c) $Cs^+$: learning (training) set - open squares; testing (validation) set - filled squares.

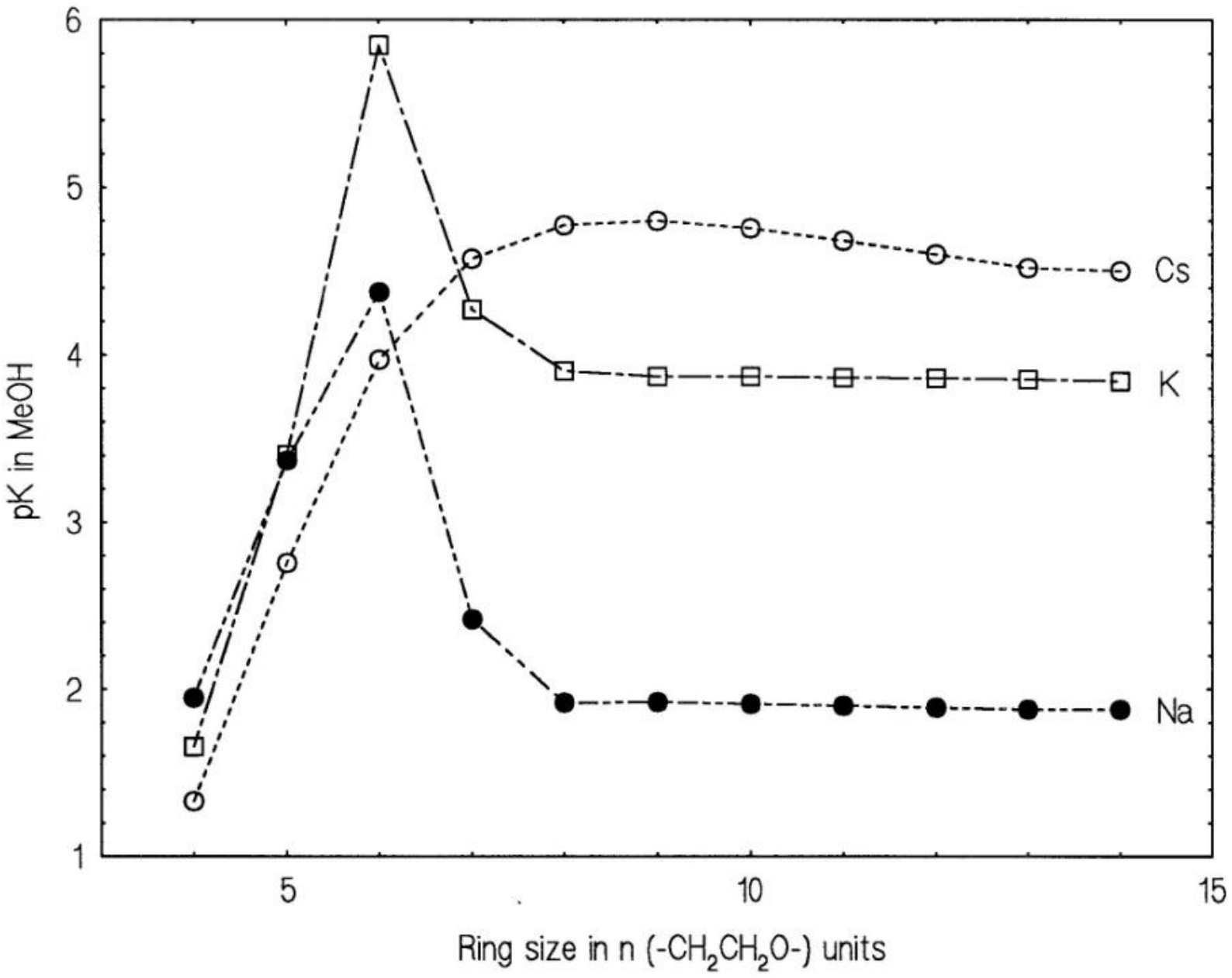

FIGURE 9 Neural network predicted values of the stability of model unsubstituted crown ethers complexes versus ring size (in $-CH_2CH_2O-$) units: a) $Na^+$: filled circles, b) $K^+$: open squares, c) $Cs^+$: open circles.

Further examination of the results produced by the neural network computational model indicates that the binding strengths reach a plateau for each of the metal ions at large ring sizes. In the case of sodium, this plateau is reached (following the maximum at $\boldsymbol{n} = 6$) for $\boldsymbol{n} = 8$ (24C8) at a value of Log K ≈ 2. Potassium also reaches a plateau at $\boldsymbol{n} = 8$, but with Log K ≈ 4. Cesium exhibits a slow steady increase of the stability constants with increasing n, reaching a plateau in the diffuse region $\boldsymbol{n} = 7\text{-}9$ with Log K ≈ 4.5; it then shows a slight decrease for larger rings.

The observation of these plateau regions suggests a systematic trend in the coordination of the 1,2-dioxyethylene units with each of the above cations. The plateau region for sodium is virtually flat at the level of Log K ≈ 2, indicating a lower overall preference of the 1,2-dioxyethylene unit for sodium as compared with potassium and cesium, where the plateau regions occur at Log K ≈ 4-5. It may be noted that this preference corresponds to the results of molecular mechanics calculations [42, 43].

**Neural Network-QSPR of Polymeric Materials [15]**

A set of molecular indices was used as input to a group of neural networks, which utilize the information to make correlations to an assortment of polymer properties. A modular-type neural network architecture, in which each property of interest is correlated to the structural representation by using individual neural networks (9 properties ⇒ 9 neural networks), was employed. This approach actually allows a better prediction of most of the properties, since each network is optimized on only one output variable, plus it allows for the fact that the same amount of data for each property is not generally available. A global prediction is, of course, possible (all properties using one network), and has been presented for some polymer properties in our previous work [15]. Table II illustrate the accuracies of using the modular neural network approach and compares the results with those obtained using a number of statistical regression techniques [linear, polynomial, and spline partial least squares regression (PLS, PPLS, SPLS), locally weighted, ridge, and kernel regression (LWR, RR, KR)] [4-9]. In all cases, cross- validation was used to determine the optimal parameters for the various forms of regression (the number of latent variables, the number of points to use in the locally weighted regression, the best values for the regression parameter theta in ridge regression, the degree of the polynomial used in polynomial and spline partial least squares, the width of the local basis functions used in KR). Most of these statistical methods assume that the data being examined can be fit to an underlying model: for PLS and RR this model is linear, for PPLS it is polynomial [the degree must be specified] and LWR is linear or quadratic. Kernel regression is general in theory, since it uses local basis sets (usually Gaussian kernels), optimized to determine the probability density function directly from the data.

**TABLE II** Predictions of various thermal and mechanical properties for polymeric materials using neural networks and statistical regression techniques*

Predicting $C_p$ [Property range: 38 – 850 J/(mol K ); 170 examples]

| Method | (%error) | Correlation | Std(residual) |
|---|---|---|---|
| NNET | 1.9 | 0.997 | 5 J/(mol K) |
| PLS | 6.0 | 0.980 | 11 J/(mol K) |
| LWR | 2.5 | 0.989 | 10 J/(mol K) |
| RR | 6.1 | 0.900 | 11 J/(mol K) |
| PPLS | 4.9 | 0.987 | 10 J/(mol K) |
| KR | 4.9 | 0.987 | 10 J/(mol K) |

Predicting $T_g$ [Property range: 130 – 685 K; 320 examples]

| Method | (%error) | Correlation | Std(residual) |
|---|---|---|---|
| NNET | 6 | 0.98 | 30° |
| PLS | 14 | 0.90 | 70° |
| LWR | 12 | 0.80 | 50° |
| RR | 14 | 0.90 | 30° |
| PPLS | 11 | 0.93 | 27° |
| KR | 8 | 0.95 | 15° |

Predicting $T_m$ [Property range: 230 – 668 K; 56 examples]

| Method | (%error) | Correlation | Std(residual) |
|---|---|---|---|
| NNET | 4 | 0.98 | 21° |
| PLS | 9 | 0.95 | 26° |
| LWR | 7 | 0.96 | 24° |
| RR | 9 | 0.95 | 26° |
| PPLS | 5 | 0.95 | 20° |
| KR | 5 | 0.95 | 40° |

Predicting $T_{de}$ [Property range: 321 – 473 K; 24 examples]

| Method | (%error) | Correlation | Std(residual) |
|---|---|---|---|
| NNET | 9 | 0.93 | 9° |
| PLS | 41 | 0.30 | 21° |
| LWR | 16 | 0.69 | 19° |
| RR | 40 | 0.23 | 25° |
| PPLS | 19 | 0.73 | 14° |
| KR | 11 | 0.82 | 15° |

Predicting Tensile Strength [Property range: 100 – 900 kg/cm$^2$; 24 examples]

| Method | (%error) | Correlation | Std(residual) |
|---|---|---|---|
| NNET | 13 | 0.90 | 40 kg/cm$^2$ |
| PLS | 34 | 0.28 | 116 kg/cm$^2$ |
| LWR | 32 | 0.25 | 205 kg/cm$^2$ |
| RR | 33 | 0.27 | 112 kg/cm$^2$ |
| PPLS | 34 | 0.31 | 86 kg/cm$^2$ |
| KR | 23 | 0.51 | 113 kg/cm$^2$ |

**TABLE II** (Continued)

Predicting Compressive Strength [Property range: 60 – 2000 kg/cm$^2$; 24 examples]

| Method | (%error) | Correlation | Std(residual) |
|---|---|---|---|
| NNET | 8 | 0.92 | 50 kg/cm$^2$ |
| PLS | 34 | 0.32 | 182 kg/cm$^2$ |
| LWR | 23 | 0.56 | 198 kg/cm$^2$ |
| RR | 31 | 0.31 | 177 kg/cm$^2$ |
| PPLS | 38 | 0.50 | 151 kg/cm$^2$ |
| KR | 17 | 0.56 | 149 kg/cm$^2$ |

Predicting Ultimate Elongation [Property range: 2 – 750%; 24 examples]

| Method | (%error) | Correlation | Std(residual) |
|---|---|---|---|
| NNET | 12 | 0.90 | 15% |
| PLS | 37 | 0.40 | 38% |
| LWR | 17 | 0.76 | 33% |
| RR | 27 | 0.63 | 39% |
| PPLS | 26 | 0.66 | 29% |
| KR | 19 | 0.62 | 51% |

Predicting Tensile Modulus [Property range: 1800 – 36000 kg/cm$^2$; 24 examples]

| Method | (%error) | Correlation | Std(residual) |
|---|---|---|---|
| NNET | 6.8 | 0.95 | 951 kg/cm$^2$ |
| PLS | 38 | 0.18 | 6000 kg/cm$^2$ |
| LWR | 25 | 0.37 | 5716 kg/cm$^2$ |
| RR | 36 | 0.14 | 5745 kg/cm$^2$ |
| PPLS | 26 | 0.48 | 3079 kg/cm$^2$ |
| KR | 19 | 0.48 | 5018 kg/cm$^2$ |

* The error measures given in Table II are defined as follows: $E_{residual}$ = abs($T+P$), where $T$ is the known answer and $P$ is the prediction; (%error) = $\Sigma(E_{residual}/T^*100)/n$, where $n$ is the number of examples; correlation = $[\Sigma(T_i - \langle T \rangle)(P_i - \langle P \rangle)]/[\sqrt{(T_i \langle T \rangle)^2}\sqrt{(O_i - \langle O \rangle)^2}]$, std (residual) = $\sqrt{[\langle E^2_{residual} \rangle - \langle E^2_{residual} \rangle / N - 1}$

Definition of the abbreviations used in the Table II:

NNET: Feedforward neural networks trained with backpropagation

PLS : Linear partial least squares regression

LWR : Locally weighted regression (a nonlinear regression technique)

RR : Ridge regression (linear technique)

PPLS : Polynomial partial least squares regression

KR : Kernel Regression

Table II indicates that the neural network prediction of the thermal and mechanical properties achieves a better accuracy than any of the statistical methods. Since the structure-property relationships are most likely nonlinear in multiple variables, it is clear that the linear models of PLS and RR will have the smallest chance of accurately predicting new results. While PPLS and LWR share the relative computational simplicity of PLS, it is possible that as a soft-modeling technique, the use of simple polynomial expansions to describe complex, nonlinear response surfaces, may still not be optimal. A more recent version of nonlinear PLS uses a spline inner relation (with user-specified knots) which, in principle, should be able to model more complex relationships. We have also used the spline-PLS (SPLS) on the problems in Table II. No notable improvement over PPLS was obtained (not shown in the Table II). Obviously, either the amount of data used was not sufficient for the statistical techniques to determine an optimal regression vector, or the correlation between the input and output is more complicated than any linear or simple polynomial functions can describe accurately (KR however should be able to fit any relationship). The power of the neural network approach is that it is model free. The best set of "functions" is determined automatically from the data, independent of the complexity of this function.

Some improvement is possible in the prediction of mechanical properties. Since the measurement of these properties is strongly dependent on the sample itself, data is usually reported with a range (lower and upper bounds). Assigning a fixed value for the mechanical properties forces the neural network to determined specific correlations to that value. A better approach would be to give the information on the bounds of the individual properties to the neural network. In this manner, even though the structural representation is constant, the correlations that are formulated by the neural network should better take into account the variability of the data. One way of representing data that have two variables per observation is to use complex domain neural networks. A neural network is given the same structure representation, but the training and the output are carried out using complex numbers. The results for the prediction of mechanical properties showed some notable improvements: average absolute error decreased by as much as 3% in certain cases.

**Molecular Design [44, 45]**

The reverse problem of property prediction, that is, the design of compounds that provide specified properties has, in the past, been addressed by: random search [46], knowledge based systems [47], graphical reconstruction [48, 49] mathematical programming [50, 51] and heuristic enumeration [52, 53]. While all of these methods can have some appeal, they suffer disadvantages, due to combinatorial complexity of the search space, design knowledge acquisition difficulties, nonlinearity in QSPR, and problems incorporating higher-level chemical knowledge. A new technique, called *computational synthesis*, which uses a hybrid of intelligent software tools, can successfully address each of these problems. This method utilizes the synergistic properties of CNNs, genetic algorithms, and fuzzy logic (methods that complement each other's strengths and compensate for disadvantages) in a modular approach to provide truly novel capabilities for computer-aided materials design. Computational synthesis involves 3 basic parts: 1) The fundamental chemical structure is transformed into N-dimensional numerical vectors by a preprocessing module; 2) CNNs (modular CNNs) are used to determine a relationship between the representation of chemical structure and a set of properties; and 3) A hybrid fuzzy logic-genetic algorithm module is used to manipulate elements into chemical structures. Modules (1) and (2) (discussed above) provide the tools for predicting properties from chemical structure and module (3) makes the reverse problem, designing chemical structures that give desired properties, possible [44, 45]. This or similar methods have proven to work very well for test compounds ranging from small organic molecules to polymeric materials [44, 45, 54-59].

Although it is clear that computational methods can provide excellent design tools capable of suggesting candidate materials suitable for a particular problem or that satisfy a set of performance criteria, in practice these methods do not formulate a single or a few candidate materials, but many possibilities are found. In any case, some ideas of the types of compounds needed to satisfy a predetermined set of properties are obtained. Additional analysis can be performed by using expert knowledge to sort out those compounds based on the feasibility of experimental synthesis (costs, complexity, etc.) or other considerations (such as health or environmental toxicology, processing requirements, etc.).

## 4. SUMMARY AND CONCLUSIONS

In summary, the use of CNNs as tools for developing complex relationships between structure and property has been found to be of great utility. A number of results in the area of QSPR have been demonstrated, which clearly point toward a concrete foundation of CNNs in this research area. CNNs can provide a method for obtaining better accuracies and more general results, and should serve as general purpose tools for QSPR in future research. Furthermore, the possibility of using these relationships to design materials that satisfy a predetermined set of properties is very promising. Hybrid and modular techniques, using CNNs, genetic algorithms, and fuzzy logic, hold much potential for providing a powerful method to meet this goal [60].

## ACKNOWLEDGEMENTS

This research was sponsored by the Divisions of Chemical Sciences and Materials Sciences, Office of Basic Energy Sciences, U.S. Department of Energy, under contract DE-AC05-840R21400 with Lockheed Martin Energy Systems, Inc; the Newly Independent States Industry Partnership Program NIS-IPP-ORS-005, Division of Defense Programs, U.S. Department of Energy, under contract DP-15 with Lockheed Martin Energy Systems, Inc; and, in part by an appointment to the ORNL Postdoctoral Research Associates Program administered jointly by Oak Ridge National Laboratory (ORNL) and the Oak Ridge Institute for Science and Education (ORISE). AAG is grateful to R. A. Sachleben and B. A. Moyer for valuable help and support.

## APPENDIX

Systematic (e.g., IUPAC) chemical names of the organic compounds represent a valuable source of structural information for generating neural network input vectors. They are unique (only one name for one given molecule, and only one structure for a given name), have distinct rules for interpretation of structural information into numeric and letter symbols, capable of classification and generalization, and have a well-developed nomenclature. These features allow easy translation of a name into a numeric vector suitable for neural network input. Training a neural network with these "chemical name - generated" vectors for known compounds allows recognition of the relationships between coded structural features of molecules and their properties. Then a trained neural network is capable of utilizing the relationships to make prediction of the properties of unknown compounds. Here we present one of the possible algorithms for the conversion of systematic chemical names into numeric vectors.

The essence of the algorithm lies in the identification of a root word, suffix, and prefix in the chemical name of a compound. The architecture of neural network (quantity of input neurons) can be derived from root word. The suffix (prefix) information is then used for translation of quantity and quality of substituents into their numeric representation.

The root word usually relates to a class of compounds, and represents the structure of a basic carbon skeleton. For example, in the name **1,3-**difluoro**benzene** the root word is **benzene**, and basic skeleton is an aromatic six-member ring. In such simple cases, determination of the number of the input neurons (six) for a representative neural network is straightforward.

The next stage is to code substituents. For example, in **1,3**-difluoro**benzene**, we have only two types of substituents: hydrogen and fluorine. The simplest way to code them is numerate them consecutively, e.g., hydrogen is 0, fluorine is 1, etc. Consecutive numeration of substituents, in most cases, gives satisfactory results. However, for better results, we recommend the use of some sort of physical characteristic of the substituent as codes, e.g., molecular weights (19 for F; 1 for H), or others. Empirical results indicate that, by choosing the value that has some physical sense for the property to be predicted, the accuracy can be improved significantly. It is also important to note that, in many systematic names, the position of hydrogen atoms are omitted, and must be reconstructed at each "unoccupied" position as default.

After coding the substituents, the input file can be generated. In our case, the numeric vector for **1,3-**difluoro**benzene** is 101000 (codes for hydrogen and fluorine are 0 and 1, respectively. If the basic skeleton has symmetry elements and, therefore, equivalent positions, all possible numeric vectors should be generated. For example **1,3-**difluoro**benzene** (101000), **2,4-**difluoro**benzene** (010100), **3,5**-difluoro**benzene** (001010), **4,6**-difluoro**benzene** (000101), **5,1**-difluoro**benzene** (100010) are the possible names of the same compound. All these numeric vectors must be included in computations; otherwise a neural network cannot determine the symmetry factor.

The names of chiral compounds require a more sophisticated input generation technique.

Stereo- isomers are usually name using S- and R prefixes, and this information should be represented in numeric codes. Thus, the input file of S-2-fluoro**hexane** can be 000 01 00 00 00 000 ($CH_3$-CH*F-$CH_2$-$CH_2$-$CH_2$-$CH_3$), but then for R-2-fluoro**hexane** it should be 000 10 00 00 00 000. (In this example 0 is a code of hydrogen, and 1 is a code of fluorine atom). If more than 2 or 3 chiral centers exist, it is recommended to use a Fisher projection for the determination of the correct numeric conversion, and follow the rules to arrange the substituents according to the IUPAC recommendations. The same situation occurs with axial- and equatorial positions in cycloalkanes.

In some limited cases the numeric vectors can be simplified. For example, input vectors for fluorochloro**propanes** contain only six numbers (the full scheme requires eight). Of those six, the first represent the number of fluorine atoms at 1 position, second represent the number of chlorine atoms at 1 position, third - the number of fluorine atoms at position 2, and so on. The numeric vectors for R- and S-1,2-dicloro-1,1,2-trifluoro**propane** are 21 11 00 and 00 11 21, respectively.

This simplified coding algorithm is also amendable for limited generalization. Thus, properties of ethanes, propanes, and butanes can be calculated using one neural network instead of three. To formulate numeric vectors of ethanes, propanes, and butanes that are compatible, zero values (00 or 000) can be added to the numeric vectors of propanes and ethanes. These zero values indicate absence of corresponding fragments in ethanes and propanes, compared with butanes.